\documentclass[lettersize,journal,10pt]{IEEEtran}
\usepackage{amsmath,amsfonts}
\usepackage{array}
\usepackage[caption=false,font=normalsize,labelfont=sf,textfont=sf]{subfig}
\usepackage{textcomp}
\usepackage{stfloats}
\usepackage{url}
\usepackage{verbatim}
\usepackage{graphicx}
\usepackage{cite}
\usepackage{amsmath,amssymb,amsfonts}
\usepackage[noend]{algorithmic}
\usepackage{algorithm}

\usepackage{stfloats}
\usepackage{textcomp}
\usepackage{enumerate}
\usepackage{xcolor}
\usepackage{booktabs}
\allowdisplaybreaks[4]

\begin{document}

\title{Permittivity Estimation in Ray-tracing Using Path Loss Data based on GAMP}

\author{Yuanhao~Jiang,  Shidong~Zhou,~\IEEEmembership{Member,~IEEE,}
and Xiaofeng~Zhong,~\IEEEmembership{Member,~IEEE}
        
\thanks{The research presented in this paper has been kindly supported by the projects as follows, 
}
\thanks{Yuanhao Jiang is with the Department of Engineering Physics, Tsinghua University, Beijing 100084, China (e-mail: jiang-yh19@mails.tsinghua.edu.cn).}
\thanks{Shidong Zhou and Xiaofeng Zhong are with the Department of Electronic Engineering, Tsinghua University, Beijing 100084, China (e-mail: zhousd@tsinghua.edu.cn; zhongxf@tsinghua.edu.cn).}
}
\markboth{Journal of \LaTeX\ Class Files,~Vol.~14, No.~8, August~2021}%
{Shell \MakeLowercase{\textit{et al.}}: A Sample Article Using IEEEtran.cls for IEEE Journals}


\maketitle
\begin{abstract}

In this paper, we propose a modified Generalized Approximate Message Passing (GAMP) algorithm to estimate permittivity parameters using path loss data in ray-tracing model. 
\end{abstract}

\section{Introduction}


The use of field wireless measurements to estimate EM parameters.

Treating the path loss data and permittivity parameters as the factor graph and using message-passing algorithms is a feasible solution. 

However, the complexity of the message-passing algorithm in \cite{JiangWCL} is exponentially related to the number of permittivity parameters. In practical scenarios with more permittivity parameters to be estimated, the computational complexity will be large. 

In this paper, we propose a modified Generalized Approximate Message Passing (GAMP) algorithm to estimate permittivity parameters using path loss data in ray-tracing model. 

\section{System Model}

\subsection{Path Loss Model}
To characterize the impact of environmental permittivity parameters on the path loss of different links, we adopt the RT model to calculate path loss data. 

Referring to the path loss model presented in \cite{JiangWCL}, which approximates the multipath effect as the superposition of energies from rays in the RT model, the measured $\hat{R}_{n}$ is derived as follows: 
\begin{subequations}
\begin{align}
\hat{R}_{n}
=& P_{n}+G_{\mathrm{tx},n}+G_{\mathrm{rx},n}+ \tilde{g}_{n}\left( \boldsymbol{\mathrm{v}}_n,\boldsymbol{\mathrm{u}}_n, \lambda, \boldsymbol{\varrho}_{e}\right) + z_{n},
\label{R} \\
\tilde{g}_{n} \triangleq&~~ \tilde{g}_{n}\left( \boldsymbol{\mathrm{v}}_n,\boldsymbol{\mathrm{u}}_n, \lambda, \boldsymbol{\varrho}_{e}\right) \notag\\ =&~~10\log_{10}{\left(\sum_{j=1}^{J} g_{j,n}\left( \boldsymbol{\mathrm{v}}_n,\boldsymbol{\mathrm{u}}_n, \lambda, \boldsymbol{\varrho}_{e}\right) \right)}.\label{tg}
\end{align}
\end{subequations}
where $P_{n}$ refers to the transmit power of the $n$-th path loss data, and $G_{\mathrm{tx},n}$, $G_{\mathrm{rx},n}$, $\lambda$ refer to the $n$-th antenna gain of the transmitters and the receivers, carrier wavelength, respectively, which are known communication conditions. 
$J$ denotes the number of rays for each path loss. 

The function $\tilde{g}_{n}(\cdot)$ denotes the gain calculation of the rays, determined by $\boldsymbol{\mathrm{v}}_n$, $\boldsymbol{\mathrm{u}}_n$, $\lambda$, and the physical environment parameters $\boldsymbol{\varrho}_{e}$. And $g_{j,n} \in \mathbb{R}$ denotes the path gain of the $j$-th ray in the linear scale. \eqref{tg} represents the energy superposition of different multipaths. 

\subsection{Permittivity Estimation Using Path
Loss Data}

In this paper, we leverage path loss data to estimate permittivity parameters which affect the wireless propagation in the physical environments. 

We consider a scenario in urban areas, where some path loss data are collected. 
These path loss data come from many sources, such as base stations and user equipments, some device-to-device links. 
The aim is to estimate permittivity parameters in the area based on these path loss data.

\begin{subequations}
\begin{align}
\text{Given}~ \epsilon_m&\in [a_m,b_m] \sim p_{\epsilon_m}(\epsilon_m)~m=1,\ldots,M\label{P1_b} \\
 (\mathrm{P}1) \quad   \hat{\boldsymbol{\epsilon}}&= \arg\max_{\boldsymbol{\epsilon}} \ln{p\left( \boldsymbol{\epsilon}\mid \mathcal{R}\right)}
    \label{p1}     
\end{align}
\end{subequations}
where
\begin{subequations}
\begin{align}
&\ln{p\left( \boldsymbol{\epsilon}\mid \mathcal{R}\right)}=\ln{p\left( \boldsymbol{\epsilon}\right)}+\sum_{\hat{R}_{n}\in\mathcal{R}} \ln{p\left( \hat{R}_{n}\mid\boldsymbol{\epsilon} \right)} + \mathrm{const}   \\
    &\ln{p\left( \hat{R}_{n}\mid\boldsymbol{\epsilon} \right)}= -\frac{\left(\hat{R}_{n}
-{R}_{n}(\boldsymbol{\epsilon})\right)^2}{2\sigma_z^2}+ \mathrm{const}
\end{align}
\end{subequations}

where $\hat{\boldsymbol{\epsilon}}$ refers to the estimated values of the permittivity parameters. 
${R}_{t,i}(\boldsymbol{\epsilon})$ denotes the calculated RSRP based on the path loss model \eqref{R}. 
$(\mathrm{P}1)$ involves estimating the permittivity modeling based on the path loss data. Due to the influence of the environment on the path loss model in \eqref{R}, the relationship between path loss data and permittivity parameters $\boldsymbol{\epsilon}$ is highly nonlinear. 

For the nonlinear estimation problem with the prior information of parameters, we consider the GAMP algorithm for solving.

\section{Modified GAMP}

In this section we present the proposed modified GAMP algorithm for estimating the relative permittivity parameters in ray-tracing models. 

The proposed modified GAMP algorithm utilizes a first-order Taylor expansion to approximate the nonlinear estimation problem as a linear mixing problem inspired by \cite{wangLowComplexityMessagePassingCooperative2017}, and then performs a multi-step GAMP with trust region. The purpose of the trust region is to ensure that the parameters to be estimated do not deviate significantly from the range of applicability of the linearization approximation.  

\subsection{Linearization}
The generalized nonlinear mixing problem is as follows:
\begin{equation}
    \hat{\boldsymbol{r}} = \boldsymbol{\tilde{g}}\left(\boldsymbol{x} \right) + \boldsymbol{z}  
\end{equation}
where $\hat{\boldsymbol{r}}\triangleq\left[\hat{R}_{1}, \ldots, \hat{R}_{N}  \right]^{\mathrm{T}}$,  $\boldsymbol{x}=\left[{x}_{1}, \ldots, {x}_{M}  \right]$ and $\boldsymbol{\tilde{g}}\left(\boldsymbol{x} \right) \triangleq \left[\tilde{g}_{1}\left(\boldsymbol{x} \right), \ldots, \tilde{g}_{N}\left(\boldsymbol{x} \right)  \right]^{\mathrm{T}}$. $\boldsymbol{z}$ denotes the $N$ additional Gaussian noise $z_i\sim \mathcal{N}(0, \sigma_z^2)$ for $i=1,\ldots,N$. 
$\hat{\boldsymbol{x}}(k)$ is denoted as the estimation of the parameters $\boldsymbol{x}$ at the $k$-th iteration, using a first-order Taylor expansion: 
\begin{equation}
    \tilde{g}_{i}\left(\boldsymbol{x} \right) \approx  \tilde{g}_{i}\left(\hat{\boldsymbol{x}}(k) \right) + \left[ \nabla \tilde{g}_{i}\left(\hat{\boldsymbol{x}}(k) \right)\right]^{\mathrm{T}}(\boldsymbol{x} - \hat{\boldsymbol{x}}(k) ) 
\end{equation}
For the sake of abbreviation, let us define: 
\begin{subequations}
\begin{align}
    \mathbf{A}(k) &= \left(\frac{\partial \tilde{g}_{i}(\boldsymbol{x})}{\partial x_m}\right)_{\substack{i=1, \ldots, N \\
m=1, \ldots, M}} \bigg|_{\boldsymbol{x}=\hat{\boldsymbol{x}}(k)} \label{A}\\
\boldsymbol{\mu}(k) &= \boldsymbol{\tilde{g}}\left(\hat{\boldsymbol{x}}(k) \right) - \mathbf{A}(k)\hat{\boldsymbol{x}}(k) \label{mu}
\end{align}
\end{subequations}

\subsection{Multi-step GAMP with Trust Region}
After linearization, a multi-step GAMP is performed. In Algorithm \ref{alg:second}, $K_\mathrm{iter}$ represents the number of iterations to perform the linearization approximation, and $K_\mathrm{gamp}$ represents the multi-step of GAMP performed in one linearization iteration. $\delta_\mathrm{tr}$ denotes the range of the trust region for each parameter. 

Here considering the additional Gaussian noise $\boldsymbol{z}_i$ and assuming a uniform priori distribution of $x_m$ within the range $[a, b]$, Steps \ref{step_GAMP_start} through \ref{step_GAMP_end} in Algorithm \ref{alg:second} provides the \textit{GAMP for MMSE estimation with AWGN output channels} \cite{ranganGeneralizedApproximateMessage2011}.
In step \ref{step_GAMP_end} $\hat{x}_m(k+1)$ is obtained through $\hat{c}_m(k),\tau_m^c(k)$:
\begin{align}
    &p\left(x_m \mid \hat{c}_m(k),\tau_m^c(k),q_{m,k_1}\right) \approx \nonumber\\ &\frac{1}{Z_\mathrm{norm}} p_{X}\left(x_m \mid q_{m,k_1}\right) \exp \left[-\frac{1}{2 \tau_m^c(k)}\left(\hat{c}_m(k)-x_m\right)^2\right]
\end{align}   

The only difference is in step \ref{step_GAMP_end}, where each $x_m$ is constrained not only by the priori distribution with the range $[a, b]$, but also by the trust region of the linearization. 
Trust region can be represented as the feasible range of approximation under the linearization when iteratively solving for $\hat{\boldsymbol{x}}(k)$.
After linearization, in order to keep  $\hat{\boldsymbol{x}}(k)$ from not converging, we consider $p_{X}\left(x_m \mid q_{m,k_1}\right)$ as both the priori distribution with the range $[a, b]$ and trust region with the range $[\hat{x}_{m}(k_1K_\mathrm{gamp})-\frac{\delta_\mathrm{tr}}{2},\hat{x}_{m}(k_1K_\mathrm{gamp})+\frac{\delta_\mathrm{tr}}{2}]$ for each $\hat{x}_{m}(k_1K_\mathrm{gamp})$. $\hat{x}_{m}(k_1K_\mathrm{gamp})$ denotes the values for performing a Taylor expansion in step \ref{Linear}. 
Hence $p_{X}\left(x_m \mid q_{m,k_1}\right)$ implies that $x_m$ follows a uniform distribution within the range $[\max{(a, \hat{x}_{m}(k_1K_\mathrm{gamp})-\frac{\delta_\mathrm{tr}}{2}}), \min{(b, \hat{x}_{m}(k_1K_\mathrm{gamp})+\frac{\delta_\mathrm{tr}}{2})}]$. 
$Z_\mathrm{norm}$ is the normalization coefficient. 

\begin{figure}[tbp]
 \begin{algorithm}[H]
	\caption{GAMP with Trust Region for Estimating Relative Permittivity of Nonlinear Problem in Ray-tracing}
	\label{alg:second}
	\begin{algorithmic}[1]
		\REQUIRE The path loss data $\hat{\boldsymbol{r}}$, the function $\boldsymbol{\tilde{g}}\left(\boldsymbol{x} \right)$, $K_\mathrm{iter}$, $K_\mathrm{gamp}$, $\delta_\mathrm{tr}$ and $\tau^w$, the inital values $\hat{\boldsymbol{x}}(0)$. 
		\ENSURE a sequence of estimates $\hat{\boldsymbol{x}}(k)$, for $k=0,1, \ldots$ through the following recursions.		
		\STATE $k_1=0$
		\WHILE{$k_1<K_\mathrm{iter}$}
        \STATE\label{Linear} Linearization: compute $\mathbf{A}(k_1K_\mathrm{gamp})$ and $\boldsymbol{\mu}(k_1K_\mathrm{gamp})$ around $\hat{\boldsymbol{x}}(k_1K_\mathrm{gamp})$ based on \eqref{A} and \eqref{mu}.
        \STATE $k_2=0$
        \WHILE{$k_2<K_\mathrm{gamp}$}
        \STATE GAMP with trust region: $k=k_1K_\mathrm{gamp}+k_2$
		\FOR{each $i$}\label{step_GAMP_start}
		\STATE 
                $\tau_i^p(k)  =\sum_m\left|a_{i m}\right|^2 \tau_m^x(k)$, \\
                $\hat{p}_i(k)  =\sum_m a_{i m} \hat{x}_m(k)-\tau_i^p(k) \hat{s}_i(k-1)$, \\
                $\hat{z}_i(k)  =\sum_m a_{i m} \hat{x}_m(k)$, \\
                ~\\
                $\hat{s}_i(k) =\frac{\hat{r}_i-\mu_i(k)-\hat{p}_i(k)}{\tau^w+\tau_i^p(k)}$, \\
$\tau_i^s(k) =\frac{1}{\tau_i^p(k)+\tau^w} $.
		\ENDFOR
		\FOR{each $m$}
		\STATE $\tau_m^c(k)  =\left[\sum_i\left|a_{i m}\right|^2 \tau_i^s(k)\right]^{-1}$, \\
$\hat{c}_m(k)  =\hat{x}_m(k)+\tau_m^c(k) \sum_i a_{i m} \hat{s}_i(k)$, \\
        ~\\
        \STATE\label{step_GAMP_end}
        $q_{m,k_1}$ is the trust region of the $k_1$-th first-order Taylor expansion, \\
        $\hat{x}_m(k+1)  =\mathbb{E}[x_m \mid \hat{c}_m(k),\tau_m^c(k),q_{m,k_1}]$, \\
$\tau_m^x(k+1)  =\operatorname{var}[x_m \mid \hat{c}_m(k),\tau_m^c(k),q_{m,k_1}]$. \label{expectation of x}
		\ENDFOR
        \ENDWHILE
		\ENDWHILE
	\end{algorithmic}
\end{algorithm}
\vspace{-0.8cm}
\end{figure}


\section{Numerical Result}

In this section, the estimation performance of the proposed algorithm is presented, using numerical path loss data. 

\begin{figure}[tbp]
	\centering
	\includegraphics[width=8.5cm]{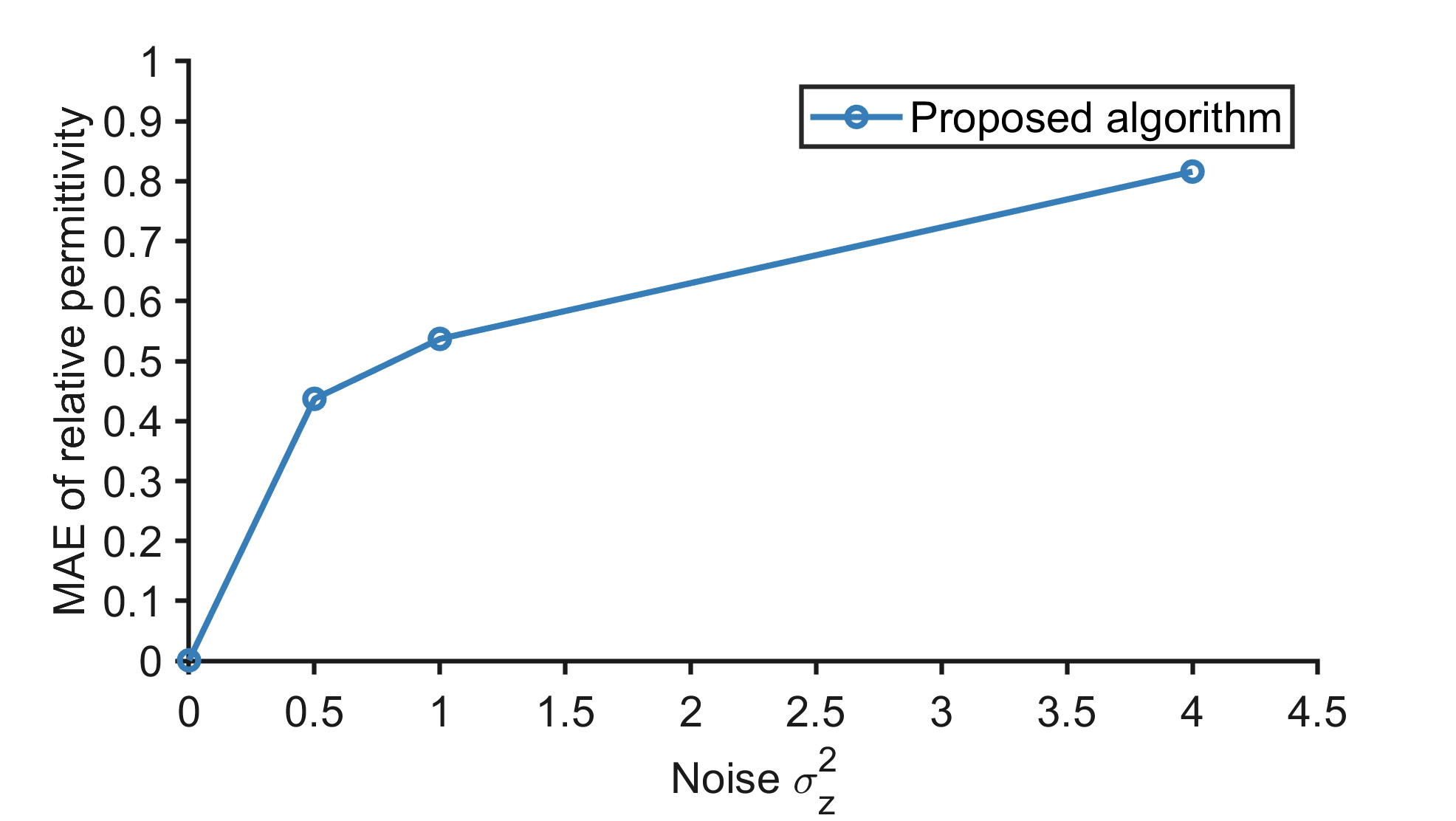}%
    \caption{Fig.~\ref{fig_sig} shows the estimation performance of the proposed algorithm under Gaussian noise with different $\sigma_z^2$.}
	\label{fig_sig}
\end{figure}

\bibliographystyle{IEEEtran}
\bibliography{ref_GAMP}

\end{document}